\documentstyle[12pt]{article}
\begin{document}
\thispagestyle{empty}
\begin{center}

\null
\vskip-1truecm
\rightline{IC/95/305,\quad  hep-th/9510067 }
\vskip1truecm
International Atomic Energy Agency\\
and\\
United Nations Educational Scientific and Cultural Organization\\
\medskip
INTERNATIONAL CENTRE FOR THEORETICAL PHYSICS\\
\vskip2truecm
{\bf VACUUM POLARIZATION\\
AND CHIRAL LATTICE FERMIONS\\}
\vskip2truecm
S. Randjbar--Daemi and J. Strathdee\\
International Centre for Theoretical Physics, Trieste, Italy.\\
\end{center}
\vskip1truecm
\centerline{ABSTRACT}
\baselineskip=18pt
\bigskip

The vacuum polarization due to chiral fermions on a 4--dimensional Euclidean
lattice is calculated according to the overlap prescription. The fermions are
coupled to weak and slowly varying background gauge and Higgs fields, and the
polarization tensor is given by second order perturbation theory. In this order
the overlap constitutes a gauge invariant regularization of the fermion vacuum
amplitude. Its low energy -- long wavelength behaviour can be computed
explicitly and we verify that it coincides with the Feynman graph result
obtainable, for example, by dimensional regularization of continuum gauge
theory. In particular, the Standard Model Callan--Symanzik RG functions are
recovered. Moreover, there are no residual lattice artefacts such as a
dependence on Wilson--type mass parameters.
\vskip2.5truecm
\begin{center}
{MIRAMARE -- TRIESTE\\
September 1995\\}
\end{center}

\newpage

\section{Introduction}

The purpose of this paper is to supplement some earlier work on the ``overlap
prescription''[1] for describing chiral fermions on a Euclidean lattice.
It was shown that this description is able to account for chiral
anomalies in the coupling of lattice fermions to a slowly varying background
gauge field and seems to give a correct formulation of the vacuum amplitude in
perturbation theory [1--3]. The notorious problem [5] of fermion ``doubling''
is
avoided in the overlap method and it has been verified that the low energy
sector is indeed chiral.

A brief analysis of vacuum polarization due to chiral fermions was given in
Ref.2. However, the discussion was incomplete in that a continuum approximation
to the overlap prescription was used without a detailed justification. The
consequent ultraviolet divergences were simply ignored. It is this aspect that
will be dealt with here. In a lattice formulation there are of course no
ultraviolet divergences. What is needed, therefore, is a somewhat more careful
discussion of how the low energy physics is to be extracted from the lattice
expressions. To illustrate how this goes, we consider here the computation of
the fermionic contribution to the vacuum polarization tensor in a chiral gauge
theory, including the effects of symmetry breaking due to a background Higgs
field. We verify that the standard results of renormalized quantum field theory
are obtained. In particular, we verify that there is no residual lattice effect
such as a dependence on the Wilson mass parameter, $r$, an effect that has been
of concern [6] to some other lattice models of chiral fermions [7].

The problem to be dealt with here, the computation of a vacuum polarization
tensor, draws on work described in previous papers [1--3]
\footnote{\normalsize See also [4] where the vacuum polarization tensor has
been discussed in a related context.}. In particular, the idea of expressing
the
vacuum amplitude for chiral fermions coupled to background fields as the
overlap
of two distinct ground states is explained in great detail in these references.
In Sec.2  we give only a brief summary of the overlap formalism in order to
explain the notation and establish that the polarization tensor is gauge
invariant even for finite lattice spacing. Some detail is provided in the
Appendices. This should be contrasted with the gauge dependent result of [8] in
the continuum 5--dimensional approach, where dimensional regularization was
used.

The general formula for the polarization tensor is developed in Sec.2. It takes
the form of a 4--dimensional Euclidean integral and represents a 1--loop
effect. The region of integration is compact, a torus -- or Brillouin zone --
of volume $(2\pi )^4$ in lattice units,
\setcounter{equation}{0}
\renewcommand{\theequation}{1.\arabic{equation}}
\begin{equation}
\Pi_{\mu\nu}(k,\Lambda ,\phi ) =\int_{BZ}\left({dp\over 2\pi}\right)^4\
F_{\mu\nu} (p,k,\Lambda ,\phi )
\end{equation}
where $\phi$ is a constant Higgs background and $\Lambda$ is an auxiliary
parameter, a kind of regulator mass that plays a central role in the overlap
formalism. The integrand, $F_{\mu\nu}$, is quite complicated and it is
fortunate that most of its detailed structure is irrelevant to the low energy
behaviour of $\Pi_{\mu\nu}$. What we need to understand is the behaviour near
$k=\Lambda =\phi =0$. To see why this is so, consider for a moment the
polarization tensor in arbitrary units where the lattice spacing, $a$, is not
equal to unity. In such units the polarization tensor, $\Pi'_{\mu\nu}$, which
has dimension 2, is given by
$$
\Pi'_{\mu\nu}(k',\Lambda ',\phi ',a)={1\over a^2}\ \Pi_{\mu\nu} (ak',a\Lambda
',a\phi ')
$$
since $k', \Lambda '$ and $\phi '$ all have dimension 1. Low energy means $k',
\Lambda '$ and $\phi '$ are small relative to the lattice scale, $a^{-1}$. We
are therefore required to consider the limit, $a\to 0$ with $k'$, etc. fixed.
Equivalently, in lattice units, we need to examine the behaviour of
$\Pi_{\mu\nu}$ in the vicinity of $k=\Lambda =\phi =0$. It turns out, as will
be discussed in Sec.3, that the integral (1.1) is singular at this point. This
is an infrared singularity, of course, since the domain of integration is
compact. In the vicinity of this point the integral is dominated by the
threshold singularities associated with the vanishing of denominators in the
expression for $F_{\mu\nu}(p,k,\Lambda ,\phi )$ due to the propagation of
chiral
fermions near their mass shell.

To compute the dominant effects, i.e. the coefficients of the singular terms in
$\Pi_{\mu\nu}$, we need to know the structure of $F_{\mu\nu}$ near $p=0$. In
this region the lattice is irrelevant and we can safely approximate the
integrand by ``continuum'' formulae, valid for $\vert p\vert\ll\pi$. We are
tacitly assuming here that there are no other threshold singularities in (1.1)
associated with massless fermion ``doublers''. It is necessary, of course, to
make sure that such doublers do not in fact appear in the low energy fermion
spectrum. The absence of such states is ensured in the overlap formalism by
keeping the parameter, $\Lambda$, large relative to $k$ and $\phi$, but small
relative to the lattice scale,
\begin{equation}
k\sim\phi\ll\Lambda\ll 1
\end{equation}

The steps needed to isolate the infrared singularity in (1.1) are explained in
some detail in Sec.3. The expression that emerges will then be recognized as
the contribution of a conventional Feynman graph of renormalized continuum
quantum field theory. In order to illustrate this equivalence we apply the
formalism to Standard Model fermions, left handed doublets and right handed
singlets, coupled to $SU(2)\times U(1)$ gauge fields and Higgs doublet
(Sec.4). We calculate the Callan--Symanzik RG functions for
quarks and leptons in the standard electroweak model.

In this paper we are exclusively concerned with the overlap approach to chiral
gauge theories on the lattice. This approach grew out of earlier work by
Kaplan [9] and others [10]. For some other approaches see [7, 11--14].

\section{The overlap prescription}

To obtained the vacuum amplitude for a collection of Weyl fermions coupled to
background gauge and Higgs fields, one begins by doubling the number of degrees
of freedom. Corresponding to each 2--component Weyl fermion introduce a
4--component Dirac field, $\psi (n)$, defined on a 4--dimensional integer
lattice, $n^\nu\in Z\!\!\!\!Z^4$. The lattice is equipped with a Euclidean
metric
tensor, $g_{\mu\nu}$, invariant with respect to one of the 4--dimensional
crystal space groups. The choice of crystal structure is probably not very
important but, in order to facilitate the emergence of $SO(4)$ invariance in
the
low energy sector, it is useful to employ a highly symmetric one such as the
$F_4$ lattice [15]. For the $F_4$ lattice one can show, for example, that the
polarization tensor is expressible in terms of two invariant functions,
\setcounter{equation}{0}
\renewcommand{\theequation}{2.\arabic{equation}}
\begin{equation}
\Pi_{\mu\nu}(k)=(k^2g_{\mu\nu}-k_\mu k_\nu )\Pi_1+g_{\mu\nu}\ \Pi_2
\end{equation}
where $k^2=g^{\mu\nu}k_\mu k_\nu$ and $\Pi_1,\Pi_2$ are functions of $k^2$ and
three other invariants of order 6, 8 and 12.

The fields $\psi (n)$ and $\psi (n)^\dagger$ satisfy anticommutation rules,
$$
\{\psi (n),\psi (m)\} =0,\quad \{\psi (n),\psi (m)^\dagger\}=\delta_{nm}1,\quad
\{\psi (n)^\dagger ,\psi (m)^\dagger \} =0
$$
and they are used in the construction of two independent Hamiltonians,
\begin{equation}
H_\pm =\sum_{n,m}\ \psi (n)^\dagger\Bigl( H(n-m)\ U(n,m\vert A)+\delta_{nm}\
M_\pm (n\vert\phi )\Bigr) \psi (m)
\end{equation}
where $H, U$ and $M_\pm$ are matrices defined as follows.

Firstly, the kinetic -- or hopping -- term, $H(n-m)$, invariant with respect to
the lattice space group is represented by the Fourier integral,
\begin{equation}
H(n-m)=\int_{BZ}\left({dp\over 2\pi}\right)^4\ e^{ip(n-m)}\
\gamma_5\Bigl( i\gamma^\mu C_\mu (p)+B(p)T_c\Bigr)
\end{equation}
where $p(n-m)=p_\mu (n-m)^\mu$ and the real functions, $C_\mu (p)$, $B(p)$ are
periodic under $p_\mu\to p_\mu +2\pi$. They are invariant with respect to the
lattice point group and are chosen such that, near $p=0$,
\begin{equation}
C_\mu (p)=p_\mu +\dots ,\qquad B(p)=rp^2+\dots
\end{equation}
Their precise form elsewhere in the Brillouin zone is not important apart from
the stipulation that they should not have a common zero anywhere except $p=0$.

The Dirac matrices are hermitian, satisfying
\begin{equation}
\{\gamma^\mu ,\gamma^\nu\} =2g^{\mu\nu},\quad \{\gamma^\mu ,\gamma_5\} =0,\quad
\gamma^2_5=1
\end{equation}
where $g_{\mu\nu}$ is the lattice metric.

The matrix $T_c$ commutes with the Dirac matrices and is diagonal with
eigenvalues $+1(-1)$ corresponding to right (left) handed Weyl fermions. It
also commutes with the gauge factor, given by the path ordered exponential,
\begin{equation}
U(n,m\vert A)=P\left(\exp i\int^n_m A\right)
\end{equation}
In this formula the path is a straight line joining lattice sites $m$ and $n$,
$$
x^\mu (t) =tn^\nu +(1-t)m^\mu ,\quad 0\leq t\leq 1
$$
and $A$ is a slowly varying vector potential,
\begin{equation}
A=dx^\mu\ A^\alpha_\mu (x)\ T_\alpha
\end{equation}

Finally, the ``mass'' term is given by the linear and gauge covariant
expression,
\begin{equation}
M_\pm (n\vert\phi )=\gamma_5 (\phi^i (n)\ T_i\pm\Lambda\  T_c)
\end{equation}
where the matrices $T_i$ incorporate Yukawa coupling parameters and specify the
representation to which $\phi^i$ belongs,
\begin{equation}
[T_i,T_\alpha ]=i(t_\alpha )_i\ ^j\ T_j
\end{equation}
They commute with Dirac matrices but they are required to anticommute with the
chirality matrix, $T_c$, since their role is to connect left handed $(T_c=-1)$
to right handed $(T_c=+1)$ fermions in order to generate mass,
\begin{equation}
\{ T_i,T_c\} =0
\end{equation}
The parameter, $\Lambda$, is a positive constant.

We shall treat the gauge field as a weak perturbation and the Higgs field as a
constant, writing
\begin{equation}
H_\pm =H^0_\pm +V
\end{equation}
where $H^0_\pm$ and $V$ can be expressed as bilinears in the Fourier
components,
$$
\psi (p) =\sum_n\ \psi (n)\ e^{-ipn}
$$
They are given as integrals over the Brillouin zone,
\begin{eqnarray}
H^0_\pm &=& \int\left({dp\over 2\pi}\right)^4\ \psi (p)^\dagger\ H_\pm (p)\psi
(p)\nonumber \\
V &=& \int\left({dp_1\over 2\pi}\right)^4\ \left({dp_2\over 2\pi}\right)^4\
\psi (p_1)^\dagger\ V(p_1,p_2)\psi (p_2)
\end{eqnarray}
where
\begin{equation}
H_\pm (p)=\gamma_5 \left( i\gamma^\mu C_\mu (p) +\phi^i T_i+(B(p)\pm\Lambda
)T_c\right)
\end{equation}
and
\begin{eqnarray}
V(p_1,p_2) &=&-\int\left({dk\over 2\pi}\right)^4\ A^\alpha_\mu (k) \
(2\pi)^4\ \delta_{2\pi}(-p_1+p_2+k)\int^1_0dt\ H(p_1-tk)^{,\mu}\
T_\alpha\nonumber\\
&&+{1\over 2}\int\left({dk_1\over 2\pi}\right)^4\left({dk_2\over
2\pi}\right)^4\ A^\alpha_\mu (k_1)\ A^\beta_\nu (k_2)\ (2\pi)^4\ \delta_{2\pi}
(-p_1+p_2+k_1+k_2)\cdot\nonumber\\
&&\cdot \int^1_0 dt_1dt_2\ H(p_1-t_1k_1-t_2k_2)^{,\mu\nu}\left(\theta
(t_1-t_2)T_\alpha T_\beta +\theta (t_2-t_1)T_\beta T_\alpha\right)\nonumber \\
&&+\dots
\end{eqnarray}
where $H(p)^{,\mu }=\partial H(p)/\partial p_\mu$, etc., and $\delta_{2\pi}$ is
the periodic delta function defined by the lattice sum,
$$
(2\pi )^4\ \delta_{2\pi}(p)=\sum_n\ e^{ipn}
$$
The momentum integrals in (2.14) are over $I\!\!R^4$ but, since $A_\mu (x)$ is
assumed to be slowly varying, its Fourier transform $A_\mu (k)$ is concentrated
around $k=0$.

The fields $\psi (n)$ and $\psi (n)^\dagger$ act on a Fock space with the
vacuum defined by $\psi (n)\vert 0\rangle =0$. To apply the overlap
prescription it is necessary to construct two distinct ground states, $\vert
A,\phi +\rangle $ and $\vert A,\phi -\rangle$ corresponding to the respective
Hamiltonians, $H_+$ and $H_-$. For free fermions, $A=\phi =0$, these ground
states $\vert +\rangle$ and $\vert -\rangle$ are obtained by filling the
negative energy eigenstates of the respective 1--body Hamiltonians, (2.13).
Modifications caused by non--vanishing $A$ and $\phi$ are then to be computed
perturbatively. The aim is to compute the functional,
\begin{equation}
e^{-\Gamma (A,\phi )}={\langle A,\phi +\vert A,\phi -\rangle\over\langle +\vert
-\rangle}\ ,
\end{equation}
to be interpreted as the vacuum amplitude. The effective potential, defined by
the ratio
\begin{equation}
e^{-\Omega V(\phi )}={\langle 0,\phi +\vert 0,\phi -\rangle\over\langle +\vert
-\rangle}\ ,
\end{equation}
where $\phi$ is constant and $\Omega$ is the volume (number of sites) of the
lattice can be computed exactly. Our main concern, however, is the polarization
tensor given by terms of second order [2] in $A$,
\begin{eqnarray}
\Gamma_{(2)}(A,\phi ) &=& {1\over 2}\ \langle 0,\phi +\vert VG^2_+V\vert 0,\phi
+\rangle+{1\over 2}\ \langle 0,\phi +\vert VG^2_-V\vert 0,\phi
-\rangle-\nonumber\\[3mm]
&&\quad {\langle 0,\phi
+\vert(VG_++G_-V+
+VG_+VG_++VG_+G_-V+G_-VG_-V)\vert 0,\phi -\rangle\over
\langle 0,\phi +\vert 0,\phi -\rangle }     \nonumber\\
&&
\end{eqnarray}
with constant $\phi $. The operators $G_\pm$ are defined by
$$
G_\pm =-{1-\vert 0,\phi\pm\rangle\langle 0,\phi\pm\vert\over H^0_\pm (\phi )}
$$
It is understood here that the free vacua are normalized zero energy
eigenstates of $H^0_\pm (\phi )$, and the Brillouin--Wigner phase convention
[1],
$\langle 0,\phi\pm\vert A,\phi\pm\rangle >0$, is used.

To evaluate the matrix elements in (2.17) it is useful to expand $\psi (p)$ in
plane wave eigenstates of the 1--body free Hamiltonians (2.13),
\begin{equation}
\psi (p)=\sum_\sigma \left( b_\pm (p,\sigma )\ u_\pm (p,\sigma )+d^\dagger_\pm
(p,\sigma )\ v_\pm (p,\sigma )\right)
\end{equation}
where $b_+$ and $d_+$ annihilate the ground state $\vert 0,\phi +\rangle$. The
orthonormal eigenspinors $u_+$ and $v_+$ satisfy
\begin{eqnarray}
H_+(p)\ u_+(p,\sigma ) &=& \omega_+(p,\sigma )\ u_+(p,\sigma )\nonumber \\
H_+(p)\ v_+(p,\sigma ) &=& -\omega_+(p,\sigma )\ v_+(p,\sigma )
\end{eqnarray}
and likewise for the eigenspinors of $H_-(p)$. The energies $\omega_\pm
(p,\sigma )$ are positive, given by
\begin{equation}
\omega_\pm (p,\sigma )^2=g^{\mu\nu}C_\mu (p)C_\nu (p)+m^2_\sigma
+(B(p)\pm\Lambda
)^2
\end{equation}
where $m^2_\sigma$ is an eigenvalue of the mass matrix, $(\phi\cdot T)^2$.
Details are given in Appendix A.

The sets $\{ u_+,v_+\}$ and $\{ u_-,v_-\}$ are both orthonormal and complete.
They are related by a unitary transformation
\begin{eqnarray}
u_- &=& \cos\beta \ u_+-\sin\beta\ v_+\nonumber\\
v_- &=& \sin\beta \ u_++\cos\beta\ v_+
\end{eqnarray}
where the angle $\beta (p,\sigma )$ can be chosen to lie in the interval
$(0,\pi /2)$. The zeroes of $\cos\beta$ serve to determine the spectrum of
fermion states in 4--dimensional spacetime. For example, the free fermion
2--point correlation function is given by
$$
{\langle 0,\phi +\vert\psi (n)\ \psi (m)^\dagger\vert 0,\phi -\rangle\over
\langle 0,\phi +\vert 0,\phi -\rangle} =\int\left({dp\over 2\pi}\right)^4\
G(p)\ e^{ip(n-m)}
$$
where
\begin{equation}
G(p)=\sum_\sigma\ {u_+(p,\sigma )\ u_-(p,\sigma )^\dagger\over \cos\beta
(p,\sigma )}
\end{equation}
In Appendix A it is shown that in the region $p,\phi\ll\Lambda\ll 1$, $\beta$
approaches the value $\pi /2$ and (2.22) reduces to the form [2]
$$
G(p)\simeq \Lambda\ {1+\gamma_5T_c\over 2}\ (ip\!\!\!/ +\phi\cdot
T)^{-1}\gamma_5 +\dots
$$
corresponding to the propagation of light chiral fermions. One can show that
there are no other light fermions if $C_\mu$ and $B$ have no common zeroes
apart from $p=0$. This is how the overlap prescription avoids the doubling
phenomenon.

Our main purpose is to study the fermionic contribution to vacuum polarization
given by (2.17) using methods sketched in Appendix A, in particular the
formula (A.12). A somewhat lengthy but straightforward calculation reduces this
to the form
\begin{eqnarray}
\Gamma_{(2)}(A,\phi ) &=& {1\over 2}\ \sum_{1,2}\
{u_+(1)^\dagger V(1,2)v_+(2)v_+(2)^\dagger V(2,1)u_+(1)\over
(\omega_+(1)+\omega_+(2))^2}\nonumber \\
&&+ {1\over 2}\ \sum_{1,2}\  {u_-(1)^\dagger V(1,2)v_-(2)v_-(2)^\dagger
V(2,1)u_-(1)\over (\omega_-(1)+\omega_-(2))^2}\nonumber \\
&&+\sum_1\left({v_+(1)^\dagger V(1,1)u_+(1)\over 2\omega_+(1)}\
\tan\beta_1-{u_-(1)^\dagger V(1,1)v_-(1)\over 2\omega_-(1)}\
\tan\beta_1\right)\nonumber \\
&&+\sum_{1,2}\ {v_+(2)^\dagger V(2,1)u_+(1)\over\omega_+(1)+\omega_+(2)}\Biggl(
-{u_+(1)^\dagger V(1,2)u_+(2)\over 2\omega_+(2)}\ \tan\beta_2
+\Biggr.\nonumber\\
&&\quad +\Biggl. {v_+(1)^\dagger V(1,2)v_+(2)\over 2\omega_+(1)}\
\tan\beta_1+{v_+(1)^\dagger V(1,2)u_+(2)\over 2(\omega_+(1)+\omega_+(2))}\
\tan\beta_1\tan\beta_2\Biggr)\nonumber\\
&&+\sum_{1,2}\ {u_-(2)^\dagger V(2,1)v_-(1)\over\omega_-(1)+\omega_-(2)}
\Biggl(-{v_-(1)^\dagger V(1,2)v_-(2)\over 2\omega_-(2)}\ \tan\beta_2 +\Biggr.
\nonumber\\
&&\quad +\Biggl. {u_-(1)^\dagger V(1,2)u_-(2)\over 2\omega_-(1)}\
\tan\beta_1+{u_-(1)^\dagger V(1,2)v_-(2)\over 2(\omega_-(1)+\omega_-(2))}\
\tan\beta_1\tan\beta_2\Biggr)\nonumber \\
&&-\sum_{1,2}\ {u_-(1)^\dagger V(1,2)v_-(2)v_+(2)^\dagger V(2,1)u_+(1)\over
(\omega_+(1)+\omega_+(2))(\omega_-(1)+\omega_-(2))}\
{1\over\cos\beta_1\cos\beta_2}
\end{eqnarray}
where the notation, $\mathop{\sum}\limits_1$, indicates integration over $p_1$
and summation over $\sigma_1$. The vertex, $V(1,2)=V(p_1,p_2)$, is given by
(2.14) as a sum of first and second order pieces. These are to be inserted
appropriately in (2.23) so as to obtain all terms of second order, i.e. the
polarization tensor.

The daunting complexity of (2.23) is mainly due to unphysical structures
associated with the extra degrees of freedom involved in the overlap
prescription. We shall be concerned with its infrared behaviour in the next
section and there it will be found that there are considerable simplifications.

Ward identities for the functional $\Gamma (A,\phi )$ are reviewed in Appendix
B. There it is shown that, in second order,
\begin{eqnarray*}
\partial_\mu\ {\delta^2\Gamma\over\delta A^\alpha_\mu (x)\ \delta A^\beta_\nu
(0)} &-& \phi^T\ t_\alpha\ {\delta^2\Gamma\over\delta\phi (x)\ \delta
A^\beta_\nu
(0)} =0\\
\partial_\mu\ {\delta^2\Gamma\over\delta A^\alpha_\mu (x)\ \delta\phi
(0)} &-& \phi^T\ t_\alpha\ {\delta^2\Gamma\over\delta\phi (x)\ \delta\phi
(0)} =\delta_4(x)\ t_\alpha\ {\delta\Gamma\over\delta\phi}
\end{eqnarray*}
where the derivatives are evaluated in the background $A_\mu =\partial_\mu\phi
=0$. Thus, it is to be expected that the polarization tensor will include a
longitudinal component if the Higgs field is non--vanishing. It represents a
fermionic contribution to the effective kinetic term for the Higgs field.

\section{The continuum limit}

The vacuum polarization tensor is defined by the second order formula (2.23)
which can be expressed in the form
\setcounter{equation}{0}
\renewcommand{\theequation}{3.\arabic{equation}}
\begin{equation}
\Gamma_{(2)}(A,\phi )={1\over 2}\int\left({dk\over 2\pi}\right)^4\ A^\alpha_\mu
(-k)\ \Pi^{\mu\nu}_{\alpha\beta}(k,\phi )\ A^\beta_\nu (k)
\end{equation}
with
\begin{equation}
\Pi^{\mu\nu}_{\alpha\beta}(k,\phi )=\int_{BZ}\left({dp\over 2\pi}\right)^4\
F^{\mu\nu}_{\alpha\beta}(p,k,\phi )
\end{equation}
Since $A_\mu (x)$ is assumed to be slowly varying, its Fourier transform $A_\mu
(k)$ is concentrated around the point $k=0$. The Higgs background, $\phi$, is
constant and, in order to identify the contributions of fermions with
physically realistic masses we shall assume that it is a small quantity, of the
same order as $k$.

At the point $k=\phi =0$ the integral (3.2) diverges. This is an infrared
singularity, a threshold effect associated with the propagation of massless
virtual fermions, and it dominates the low energy structure of the polarization
tensor. Our task is to isolate this singularity and compute its contribution
for small values of $k$ and $\phi$.

In the analytic continuation of (3.2) to complex values of $k$ other,
unphysical, singularities are encountered. For example, if $\Lambda\ll 1$,
there is a branch point at $k^2=-4\Lambda^2$ that can be interpreted as the
threshold for creation of a pair of fermions of mass, $\Lambda$. Such
thresholds are artefacts of the overlap prescription and have no physical
significance. The overlap method can be expected to give a physically
trustworthy result only in the restricted region,
\begin{equation}
k\sim\phi\ll\Lambda\ll 1
\end{equation}

The infrared singularity, being a threshold effect, must arise from the
simultaneous vanishing of factors in the denominator of the integrand of (3.2).
Since the momentum integral is 4--dimensional, it is necessary to have a zero
of order 4 in the denominator in order to create an infrared divergence at
$k=\phi =0$. Generically, there must be four factors, each contributing a
simple zero. It is quite easy to identify the responsible terms and it will be
seen that their zeroes occur at $p=0$ and nowhere else.

To identify the sources of vanishing denominators in (2.23) we need the
explicit formulae for eigenspinors developed in Appendix A,
\begin{eqnarray}
u_\pm (p,\sigma ) &=& {\omega_\pm +H_\pm (p)\over
\sqrt{2\omega_\pm (\omega_\pm +B\pm\Lambda )}}\ \chi (\sigma )\nonumber \\
v_\pm (p,\sigma ) &=& {\omega_\pm -H_\pm (p)\over
\sqrt{2\omega_\pm (\omega_\pm -B\mp\Lambda )}}\ \chi (\sigma )
\end{eqnarray}
where $H_\pm (p)$ are the 1--body Hamiltonians (2.13) and the positive energies
$\omega_\pm$ are given by (2.20). Also,
\begin{eqnarray}
\cos\beta &=& v^\dagger_+v_-\nonumber\\
&=&\sqrt{{\omega_+-B-\Lambda\over 2\omega_+}\ {\omega_--B+\Lambda\over
2\omega_-}}+
\sqrt{{\omega_++B+\Lambda\over 2\omega_+}\ {\omega_-+B-\Lambda\over 2\omega_-}}
\end{eqnarray}
An inspection of the various terms in (2.23) shows that $\cos\beta$ must
contribute a zero if we are to obtain a zero of order 4. In order for
$\cos\beta$ to vanish, {\em both} non--negative square roots in (3.5) must
vanish. There are two possibilities, either
\begin{equation}
\omega_+=B+\Lambda,\qquad \omega_-=-B+\Lambda
\end{equation}
or else,
$$
\omega_+=-B-\Lambda,\qquad \omega_-=B-\Lambda   \eqno(3.6')
$$
The second alternative can be excluded immediately because (3.6$'$) implies
$\omega_++\omega_-=-2\Lambda$ which contradicts the positivity of $\omega_\pm$.
The first alternative is acceptable only if $\Lambda\pm B>0$, i.e. if
$\Lambda^2>B^2$. On the other hand (3.6) implies
$C_\mu =\phi =0$ as can be seen from the definition (2.20). Recall now, that it
was part of the overlap prescription that $C_\mu$ and $B$ should have no common
zeroes except for the origin, $p=0$. In particular, this means that $B$ should
not vanish at any of the non--trivial zeroes of $C_\mu$. Hence, by choosing
$\Lambda$ small enough we can ensure that the requirement $\Lambda\pm B>0$ is
satisfied only at the origin where $B$ vanishes. We conclude that, if $\Lambda$
is small enough then the only zero of $\cos\beta$ is at $p=0$. At this point
(3.6) also implies zeroes in the denominators of $u_-$ and $v_+$. Hence, there
is an infrared singularity and it resides in the terms,
\begin{eqnarray}
\Gamma_{sing}(A,\phi ) &=& \sum_{1,2}\ \Biggl[ {v_+(2)^\dagger
V(2,1)u_+(1)v_+(1)^\dagger V(1,2)u_+(2)\over 2(\omega_+(1)+\omega_+(2))^2}\
\tan\beta_1\tan\beta_2\Biggr.\nonumber \\
&+&{u_-(2)^\dagger V(2,1)v_-(1)u_-(1)^\dagger V(1,2)v_-(2)\over
2(\omega_-(1)+\omega_-(2))^2}\
\tan\beta_1\tan\beta_2\nonumber \\
&-&\Biggl. {v_+(2)^\dagger V(2,1)u_+(1)u_-(1)^\dagger V(1,2)v_-(2)\over
(\omega_+(1)+\omega_+(2))(\omega_-(1)+\omega_-(2))}\
\cos\beta_1\cos\beta_2
\end{eqnarray}
To identify the singularity we can expand around the point $p=k=\phi =0$ using
\begin{eqnarray*}
\omega_\pm &=& \Lambda +{p^2+m^2_\sigma\over 2\Lambda}\ \pm\ rp^2 +\dots\\
\cos\beta &=& \sqrt{{p^2+m^2_\sigma\over \Lambda^2}}\ +\dots\\
V(1,2) &=& -i\gamma_5\gamma^\mu\ A^\alpha_\mu (p_1-p_2)\ T_\alpha +\dots
\end{eqnarray*}
etc. Using the approximate expressions for the eigenspinors given in Appendix A
one obtains the simple result,
\begin{equation}
\Pi^{\mu\nu}_{\alpha\beta} (k,\phi ) = -\int\left({dp\over 2\pi}\right)^4\
{\rm Tr}\Biggl[\gamma^\mu T_\alpha\ {1+\gamma_5 T_c\over 2}\ {1\over
ip\!\!\!/_1+\phi\cdot T}
\cdot\gamma^\nu T_\beta\ {1+\gamma_5T_c\over 2}\ {\over ip\!\!\!/_2
+\phi\cdot T} +\dots\Biggr]
\end{equation}
where $p_1=p+\displaystyle{k\over 2},\ \  p_2=p-{k\over 2}$. Terms not
shown here
do not contribute to the infrared singularity.

To compute the singular terms contributed by (3.8) one can apply standard field
theory methods. For example, using a Laplace transform representation for the
vanishing denominator factors,
$$
{1\over p^2_1+m^2_1}\ {1\over p^2_2+m^2_2}=\int^1_0 dx\int^\infty_0dt\ t\
e^{-t[x(p^2_1+m^2_1)+(1-x)(p^2_2+m^2_2)]}
$$
the integral can be expressed in the form
\begin{equation}
\Pi^{\mu\nu}_{\alpha\beta}(k,\phi )=\int^1_0dx\int^\infty_0{dt\over t}\
f^{\mu\nu}_{\alpha\beta}(t,x,k,\phi )\ e^{-tx(1-x)k^2}
\end{equation}
where
\begin{eqnarray}
&&f^{\mu\nu}_{\alpha\beta}(t,x,k,\phi ) =-t^2\int\left({dp\over 2\pi}\right)^4\
e^{-t[xp^2_1+(1-x)p^2_2-x(1-x)k^2]}\nonumber\\
&&\cdot {\rm Tr}\left[\gamma^\mu T_\alpha{1+\gamma_5 T_c\over
2}(-ip\!\!\!/_1+\phi\cdot T) e^{-tx(\phi T)^2}\ \gamma^\nu T_\beta {1+\gamma_5
T_c\over 2} (-ip\!\!\!/_2+\phi\cdot T) e^{-t(1-x)(\phi T)^2}+\dots\right]
\nonumber\\
&&
\end{eqnarray}
The behaviour at small $k^2$ of the integral (3.9) is dominated by the
behaviour of $f$ at large $t$ and this can be determined from (3.10) by the
method of steepest descents. Indeed it is clear that the factor $\exp (-tp^2)$
in the integrand of (3.10) serves to concentrate the support around $p=0$ as
$t$ becomes large. This validates the expansion used in obtaining (3.8). The
leading term in an asymptotic expansion in $1/t$ is given by
\begin{equation}
f^{\mu\nu}_{\alpha\beta}\simeq {-2\over (4\pi )^2}\left[\left\{
-2x(1-x)(k^2g^{\mu\nu}-k^\nu k^\nu )+g^{\mu\nu}\left( x(1-x)k^2+{1\over
t}\right)\right\}
\rho_{\alpha\beta}+g^{\mu\nu}\sigma_{\alpha\beta}+\dots\right]
\end{equation}
where $\rho_{\alpha\beta}$ and $\sigma_{\alpha\beta}$ are defined by the
flavour traces,
\begin{eqnarray}
\rho_{\alpha\beta} &=& {\rm tr}\left( T_\alpha\ e^{-tx(\phi T)^2}\ T_\beta\
e^{-t(1-x)(\phi T)^2}\right)\nonumber\\
\sigma_{\alpha\beta} &=& {\rm tr}\left( T_\alpha\ \phi\cdot T\ e^{-tx(\phi
\cdot T)^2}\ T_\beta\
\phi\cdot T\ e^{-t(1-x)(\phi\cdot T)^2}\right)
\end{eqnarray}
On substituting (3.11) into (3.9) one obtains the asymptotic formula
\begin{eqnarray}
\Pi^{\mu\nu}_{\alpha\beta} (k,\phi ) &\simeq& -{2\over (4\pi
)^2}\int^1_0dx\int^\infty {dt\over t}\Biggl[ -2x(1-x)\
\rho_{\alpha\beta}(t,\phi )(k^2g^{\mu\nu}-k^\mu k^\nu )+\Biggr.\nonumber \\
&&\Biggl. +\left( \sigma_{\alpha\beta}(t,\phi )+{\partial\over\partial t}\
\rho_{\alpha\beta}(t,\phi )\right) g^{\mu\nu}\Biggr] e^{-tx(1-x)k^2}
\end{eqnarray}
valid for $k,\phi\to 0$. The integral over $t$ can
be written as a linear combination of terms like
\begin{equation}
\int^\infty {dt\over t}\ e^{-tM^2} =-\ell n\ M^2+c_0+c_1M^2+\dots
\end{equation}
for $M^2\to 0$. The leading term is well defined but the subleading ones are
sensitive to what happens at finite values of $t$ where the expression (3.11)
is not adequate.

In Sec.4 a specific example drawn from the standard model will be considered in
more detail. To conclude the present discussion we remark only that the flavour
traces (3.12) take the form
\begin{equation}
\rho_{\alpha\beta}=\sum\ \rho^i_{\alpha\beta}\ e^{-tm^2_i},\quad
\sigma_{\alpha\beta} =\sum\ \sigma^i_{\alpha\beta}\ e^{-tm^2_i}
\end{equation}
where $m_i$ is a fermion mass. Substituting these expressions into (3.13) and
applying (3.14) one obtains
\begin{eqnarray}
\Pi^{\mu\nu}_{\alpha\beta} (k,\phi ) &\simeq& -{2\over (4\pi )^2}\int^1_0 dx
\Bigl[ 2x(1-x)\ \rho^i_{\alpha\beta}(k^2g^{\mu\nu}-k^\mu k^\nu
)-\Bigr.\nonumber \\
&&\Bigl. -(\sigma^i_{\alpha\beta} -m^2_i\
\rho^i_{\alpha\beta})g^{\mu\nu}\Bigr]\
\ell n (m^2_i+x(1-x)k^2)
\end{eqnarray}
valid for $m,k\ll\Lambda\ll 1$.

\section{Standard model}

To study the application of the overlap prescription to the standard model we
consider the continuum limit of Sec.3 for the gauge group $SU(2)\times U(1)$
with a single Higgs doublet and four chiral fermions, a left handed doublet and
two right handed singlets,
$$
\phi =\left(\matrix{ u_L\cr d_L\cr u_R\cr d_R\cr}\right)
$$
The generators of $SU(2)\times U(1)$ are represented by the $4\times 4$
matrices,
\setcounter{equation}{0}
\renewcommand{\theequation}{4.\arabic{equation}}
\begin{eqnarray}
T_a&=&\left(\matrix{ {1\over 2}\tau_a&&\cr &0&\cr &&0\cr}\right)\ ,\quad
a=1,2,3\\ [3mm]
T_Y&=&\left(\matrix{ Y_L&&\cr &Y_u&\cr &&Y_d\cr}\right)
\end{eqnarray}
where the hypercharge assignments are given by
\begin{equation}
(Y_L,Y_u,Y_d)=\left\{\matrix{
(1/3, 4/3, -2/3),\quad &{\rm quarks}\hfill\cr &\cr
(-1, 0, -2), &{\rm leptons}\cr}\right.
\end{equation}
(The Higgs doublet carries $Y_\phi =1$.) The chirality matrix is
\begin{equation}
T_c=\left(\matrix{ -1\!\!1&&\cr &1&\cr &&1\cr}\right)
\end{equation}
and the Higgs--Yukawa matrix is
\begin{equation}
\phi\cdot T =\left(\matrix{
0&f_u\bar\phi &f_d\phi\cr
f_u\bar\phi^\dagger &0&0\cr
f_d\phi^\dagger&0&0\cr}\right)
\end{equation}
where $f_u$ and $f_d$ are real Yukawa coupling constants and
\begin{equation}
\phi =\left(\matrix{ \phi_1\cr \phi_2\cr}\right),\quad \bar\phi =
\left(\matrix{ \phi^*_2\cr -\phi^*_1\cr}\right)
\end{equation}
The matrices $T_a, T_c$ and $\phi\!\cdot\! T$ are all hermitian.

It is a simple exercise to compute the flavour traces $\rho_{\alpha\beta}$ and
$\sigma_{\alpha\beta}$ defined by (3.12). One finds
\begin{eqnarray}
\rho_{ab} &=& {1\over 4}\ \delta_{ab} \left (e^{-t(xm^2_u+(1-x)m^2_d)} +
e^{-t(xm^2_d+(1-x)m^2_u)}\right)\nonumber\\
&&+{1\over 4}\ {\phi^\dagger\sigma_a\phi\phi^\dagger\sigma_b\phi\over
(\phi^\dagger\phi )^2}\Biggl[ e^{-tm^2_u}+e^{-tm^2_d}\Biggr.\nonumber\\
&&\Biggl.\quad - e^{-t(xm^2_u+(1-x)m^2_d)}-e^{-t(xm^2_d+(1-x)m^2_u)}
\Biggr]\nonumber\\
\rho_{aY} &=& {Y_L\over 2}\ {\phi^\dagger\tau_a\phi\over\phi^\dagger\phi}\
(-e^{-tm^2_u}+e^{-tm^2_d})\nonumber\\
\rho_{YY} &=& (Y^2_L+Y^2_u)\ e^{-tm^2_u}+(Y^2_L+Y^2_d)\ e^{-tm^2_d}\nonumber \\
\sigma_{ab} &=& 0\nonumber \\
\sigma_{aY} &=& {1\over 2}\ {\phi^\dagger\tau_a\phi\over \phi^\dagger\phi}\
(-Y_um^2_ue^{-tm^2_u}+Y_dm^2_de^{-tm^2_d})\nonumber\\
\sigma_{YY} &=& 2Y_LY_um^2_ue^{-tm^2_u}+2Y_LY_dm^2_de^{-tm^2_d}
\end{eqnarray}
where $m^2_u=f^2_u\phi^\dagger\phi$ and $m^2_d=f^2_d\phi^\dagger\phi$.
In writing $\rho_{ab}$ we dropped a term which is antisymmetric with respect
to $x\leftrightarrow 1-x$. The coefficients $\rho^i_{\alpha\beta}$ and
$\sigma^i_{\alpha\beta}$ of (3.15) are thereby determined and can be
substituted
into (3.16). We list the transverse and longitudinal components separately,
writing
\begin{equation}
\Pi^{\mu\nu}_{\alpha\beta}(k,\phi )=(k^2g^{\mu\nu}-k^\mu k^\nu
)\Pi_{\alpha\beta} + g^{\mu\nu}\ H_{\alpha\beta}
\end{equation}
The results are
\begin{eqnarray}
\Pi_{ab} &=& {1\over 16\pi^2}\int^1_0dx\ x(1-x)\Biggl[ -2\delta_{ab}\ \ell n
(xm^2_u+(1-x)m^2_d+x(1-x)k^2)\Biggr.\nonumber\\
&&+{\phi^\dagger\tau_a\phi\phi^\dagger\tau_b\phi\over (\phi^\dagger\phi )^2}
\Bigl\{ -\ell n (m^2_u+x(1-x)k^2)-\ell n(m^2_d+x(1-x)k^2)\Bigr.\nonumber\\
&&\qquad \Biggl.\Bigl. +2\ell n
(xm^2_u+(1-x)m^2_d-x(1-x)k^2\Bigr\}\Biggr]\nonumber\\
&&\nonumber\\
\Pi_{aY} &=& {Y_L\over 8\pi^2}\
{\phi^\dagger\tau_a\phi\over\phi^\dagger\phi}\int^1_0dx\ x(1-x)\ \ell n\left(
{m^2_u+x(1-x)k^2\over m^2_d+x(1-x)k^2}\right)\nonumber\\
&&\nonumber\\
\Pi_{YY} &=&-{1\over 4\pi^2}\int^1_0dx\ x(1-x)\Biggl[ (Y^2_L+Y^2_u)\ \ell n
(m^2_u+x(1-x)k^2)\Biggr.\nonumber\\
&&\hspace{5cm} \Biggl. +(Y^2_L-Y^2_d)\ \ell n
(m^2_d+x(1-x)k^2)\Biggr]\nonumber\\
&&\nonumber\\
H_{ab} &=& {1\over 16\pi^2}\int^1_0dx\Biggl[ -\delta_{ab} (xm^2_u+(1-x)m^2_d)\
\ell n(xm^2_u+(1-x)m^2_d+x(1-x)k^2)\Biggr.\nonumber\\
&&\nonumber\\
&&-{1\over 2}\ {\phi^\dagger\tau_a\phi^\dagger\tau_b\phi\over
(\phi^\dagger\phi )^2}\ \Bigl\{ m^2_u\ \ell n(m^2_u+x(1-x)k^2)+m^2_d\ \ell
n(m^2_d+x(1-x)k^2)\Bigr.\nonumber\\
&&\qquad \Biggl.\Bigl. -2(xm^2_u+(1-x)m^2_d)\ \ell n
(xm^2_u+(1-x)m^2_d+x(1-x)k^2)\Bigr\}\Biggr]\nonumber\\
&&\nonumber\\
H_{aY} &=& {1\over 16\pi^2}\ {\phi^\dagger\tau_a\phi\over\phi^\dagger\phi}
\int^1_0dx\Biggl[ (Y_L-Y_u)m^2_u\ \ell n (m^2_u+x(1-x)k^2)-\Biggr.\nonumber\\
&&\hspace{5cm} \Biggl. - (Y_L-Y_d)m^2_d\ \ell
n(m^2_d+x(1-x)k^2)\Biggr]\nonumber\\ &&\nonumber\\
H_{YY} &=& -{1\over 8\pi^2}\int^1_0dx\Biggl[ (Y_L-Y_u)^2m^2_u\ \ell
n(m^2_u+x(1-x)k^2)+\Biggr.\nonumber\\
&&\hspace{5cm}\Biggl. +(Y_L-Y_d)^2m^2_d\ \ell n(m^2_u+x(1-x)k^2)\Biggr]
\end{eqnarray}
These expressions comprise the leading, or infrared singular, part of the
unrenormalized polarization tensor. The first subleading terms, corresponding
to the parameters, $c_0$, in (3.14) would take the form of a second order
polynomial in $k$. Because gauge invariance is guaranteed, this polynomial
whose coefficients must depend on $\Lambda$ as well as details of the functions
$C_\mu$ and $B$, can be absorbed in a counterterm. Higher order effects,
involving $c_1,\dots$ become vanishingly small in the infrared limit and can be
discarded. Therefore, the only remaining dependence on the lattice
regularization is in the system of units we have employed. This also can be
absorbed in a counterterm.

Define the renormalized polarization tensor by introducing a reference mass,
$\mu$, and using it to scale the arguments of all the logarithms in the
expressions (4.9), i.e. make the replacements, $k^2\to k^2/\mu^2, m^2\to
m^2/\mu^2$. This generates a counterterm proportional to $\ell n\ \mu^2$,
\begin{equation}
\Gamma =\Gamma^R+\Delta\Gamma
\end{equation}
where the counterterm is given by the gauge invariant local expression
\begin{eqnarray}
\Delta\Gamma &=& -\ell n\ \mu^2\int d^4x\Biggl[ {1\over 48\pi^2}\ {1\over 4}\
(F^a_{\mu\nu})^2+{2Y^2_L+Y^2_u+Y^2_d\over 24\pi^2}\ {1\over 4}\
(F^Y_{\mu\nu})^2+\Biggr.\nonumber\\
&&\hspace{6cm} +\Biggl. {f^2_u+f^2_d\over 16\pi^2}\
\nabla_\mu\phi^\dagger\nabla_\mu\phi\Biggr]
\end{eqnarray}
It can be combined with corresponding terms in the classical action to define
running coupling constants
\begin{eqnarray}
{1\over g^2_I} &=& {1\over g^2_{0I}} -{1\over 48\pi^2}\ \ell n\
\mu^2\nonumber\\
{1\over g^2_Y} &=& {1\over g^2_{0Y}}-{2Y^2_L+Y^2_u+Y^2_d\over 24\pi^2}\ \ell n\
\mu^2
\end{eqnarray}
and the wave function renormalization
\begin{equation}
Z_\phi =1-{f^2_u+f^2_d\over 16\pi^2}\ \ell n\ \mu^2
\end{equation}
The fermionic contributions to the Callan--Symanzik functions are therefore
given by
\begin{eqnarray}
\mu\ {\partial g_I\over\partial\mu} &=& {1\over 48\pi^2}\ g^3_I\nonumber\\
\mu\ {\partial g_Y\over \partial\mu} &=& {2Y^2_L+Y^2_u+Y^2_d\over 24\pi^2}\
g^3_Y\nonumber\\[2mm]
&=&\left\{\matrix{ {22/9\over 24\pi^2}\ g^3_Y, &{\rm quarks}\hfill\cr &\cr
{6\over 24\pi^2}\ g^3_Y, &{\rm leptons}\cr}\right.\nonumber\\[2mm]
\mu\ {\partial\ell n\ Z^{-1}_\phi\over\partial\mu}
&=& {f^2_u+f^2_d\over 8\pi^2}
\end{eqnarray}

\section{Summary}

Our purpose in this work was to show that a lattice formulation of gauge
theories with chiral fermions is satisfactory in the lowest order of
perturbation theory. We have computed the vacuum polarization due to fermions
and verified that it reduces in the low energy approximation -- or,
equivalently, when the lattice spacing goes to zero -- to the expected form. In
obtaining this result it was crucial that the free fermion spectrum contains no
light doublers; a feature that can be ensured by appropriate choice of the free
Hamiltonians.

The possibility to choose Hamiltonians such that light fermions are not doubled
seems to be the real strength of the overlap prescription. It should be
emphasized that this has nothing to do with the Higgs mechanism. We introduced
a Higgs multiplet into the system only in order to make it phenomenologically
more interesting. This field played no role in solving the doubling problem.

Our approach in this work has been to emphasize the infrared nature of the
problem by scaling with respect to the ultraviolet cutoff, i.e. working in
lattice spacing units. Integrals that are not infrared dominated can be set
aside and absorbed in the usual counterterms. On the other hand, integrals that
are infrared dominated become singular when the masses and external momenta are
scaled to zero, and these singular contributions can be computed without regard
to short distance structure. They necessarily coincide with what is usually
called ``the finite part'' in Feynman graph calculations. As it happens, the
amplitude treated here is gauge invariant and hence so are the counterterms.
But this is not important. Evaluation of counterterms necessarily involves
integrating over the entire Brillouin zone and this would be a difficult task.

Of course, this is only a second order calculation. In previous work it was
shown that the expected chiral anomalies begin to emerge in the next order, and
that is also satisfactory. However, not much is known about the general
structure of the overlap amplitude, apart from several numerical studies [1],
[16] which support the viability of the formalism.

\section*{Acknowledgments}

We are appreciative of useful conversations with H. Arfaei and with
H. Schlereth who brought Ref.[6]
to our attention. Fruitful correspondence with H. Neuberger is also
acknowledged.

\newpage

\section*{Appendix A -- Free fermion states}

To compute overlaps and matrix elements for free fermions $(A=0,\phi =$
constant) it is necessary first to diagonalize the Hamiltonians $H^0_\pm$ and
construct their respective ground states. To this end consider the 1--body
operators given by (2.9a)
\setcounter{equation}{0}
\renewcommand{\theequation}{A.\arabic{equation}}
\begin{equation}
H_\pm (p) =\gamma_5(i\gamma^\mu\ C_\mu (p)+\phi^i T_i+(B(p)\pm\Lambda )T_c)
\end{equation}
The square of this matrix,
$$
H_\pm (p)^2=g^{\mu\nu} C_\mu C_\nu +(\phi\cdot T)^2+(B\pm\Lambda )^2
$$
is free of Dirac matrices and we can diagonalize it by choosing momentum
independent orthonormal spinors $\chi (\sigma )$ such that
\begin{equation}
(\phi\cdot T)^2\ \chi (\sigma ) =m^2_\sigma\chi (\sigma )
\end{equation}
It is convenient also to impose the condition
\begin{equation}
\gamma_5 T_c\ \chi (\sigma )=\chi (\sigma )
\end{equation}
Orthonormal eigenspinors of (A.1) can then be written,
\begin{eqnarray}
u_\pm (p,\sigma ) &=& {\omega_\pm +H_\pm (p)\over \sqrt{2\omega_\pm (\omega_\pm
+B\pm\Lambda )}}\ \chi (\sigma )\nonumber\\
v_\pm (p,\sigma ) &=& {\omega_\pm -H_\pm (p)\over \sqrt{2\omega_\pm (\omega_\pm
-B\mp\Lambda )}}\ \chi (\sigma )
\end{eqnarray}
where the eigenvalues are given by
\begin{equation}
\omega_\pm (p,\sigma )=\sqrt{g^{\mu\nu}C_\mu C_\nu +m^2_\sigma +(B\pm\Lambda
)^2}
\end{equation}
These numbers are chosen to be positive so that $u_\pm$ and $v_\pm$ are,
respectively, positive and negative energy eigenspinors of $H_\pm (p)$. The
sets $\{ u_+,v_+\}$ and $\{ u_-,v_-\}$ are both orthonormal and complete.
They are related by a unitary transformation,
\begin{eqnarray}
u_- &=& \cos\beta\ u_+-\sin\beta\ v_+\nonumber \\
v_- &=& \sin\beta\ u_++\cos\beta\ v_+
\end{eqnarray}
where the angle $\beta =\beta (p,\sigma )$ lies in the range $(-\pi /2,\pi
/2)$,
\begin{eqnarray}
\cos\beta &=& \sqrt{{\omega_+-B-\Lambda\over 2\omega_+}\
{\omega_--B+\Lambda\over 2\omega_-}} +
\sqrt{{\omega_++B+\Lambda\over 2\omega_+}\ {\omega_-+B-\Lambda\over
2\omega_-}}\nonumber \\
\sin\beta &=& \sqrt{{\omega_++B+\Lambda\over 2\omega_+}\
{\omega_--B+\Lambda\over 2\omega_-}} -
\sqrt{{\omega_+-B-\Lambda\over 2\omega_+}\
{\omega_-+B-\Lambda\over 2\omega_-}}
\end{eqnarray}

Creation and annihilation operators for 1--fermion states are defined by the
plane wave expansions,
$$
\psi (n) =\int\left({dp\over 2\pi}\right)^4\ e^{ipn}\ \sum_\sigma
\left( b_\pm (p,\sigma )u_\pm (p,\sigma )+d^\dagger_\pm (p,\sigma )v_\pm
(p,\sigma )\right)  \eqno({\rm A}.7')
$$
They are related by a Bogoliubov transformation,
\begin{eqnarray}
b_- &=& \cos\beta\ b_+-\sin\beta\ d^\dagger_+\nonumber \\
d^\dagger_- &=& \sin\beta\ b_++\cos\beta\ d^\dagger_+
\end{eqnarray}

The Dirac vacua are defined by filling the negative energy states,
$$
\vert\pm\rangle =\mathop{\Pi}\limits_{k,\sigma}\ d_\pm (k,\sigma )\vert
0\rangle
\eqno({\rm A}.8')
$$
and their overlap is easily obtained,
\begin{equation}
\langle +\vert -\rangle = \mathop{\Pi}\limits_{k,\sigma} \cos\beta (k,\sigma )
\end{equation}
It is now straightforward to construct correlation functions. For example, the
2--point function,
\begin{eqnarray}
\langle +\vert\psi (n)\psi^\dagger (m)\vert -\rangle\over\langle +\vert
-\rangle &=&
\int\left({dp\over 2\pi}\right)^4\left({dp'\over
2\pi}\right)^4\sum_{\sigma\sigma '}\ u_+(p,\sigma )\ {\langle +\vert
b_+(p,\sigma )b^\dagger_-(p',\sigma ')\vert -\rangle\over\langle +\vert
-\rangle} \ u^\dagger_-(p',\sigma ')\nonumber\\
&=& \int_{BZ} \left({dp\over 2\pi}\right)^4\ \sum_\sigma\ {u_+(p,\sigma
)u^\dagger_-(p,\sigma )\over \cos\beta (p,\sigma )}
\end{eqnarray}
It is instructive to examine the long distance behaviour of this correlation
function. For $p\sim\phi\ll\Lambda\ll 1$, the energies (A.5) are approximated
by
$$
\omega_\pm (p,\sigma )\simeq\Lambda +{p^2+m^2_\sigma\over 2\Lambda}\ \pm\
rp^2+\dots
$$
The eigenspinors (A.4) reduce to the form
$$
u_+(p,\sigma )\simeq\chi (\sigma ),\quad u_-(p,\sigma )\simeq \gamma_5\
{ip\!\!\!/ +\phi\cdot T\over\sqrt{p^2+m^2_\sigma}}\ \chi
$$
and
$$
\cos\beta (p,\sigma )\simeq \sqrt{{p^2+m^2_\sigma\over\Lambda^2}}+\dots
$$
Hence, in this neighbourhood,
\begin{equation}
\sum_\sigma\ {u_+(p,\sigma )u^\dagger_-(p,\sigma )\over\cos\beta (p,\sigma )}
\simeq \Lambda\ {1+\gamma_5T_c\over 2}\ {1\over ip\!\!\!/ +\phi\cdot T}\
\gamma_5+\dots
\end{equation}
and the correlator (A.10) approximates -- apart from the factors $\Lambda$ and
$\gamma_5$, which can be removed by a suitable redefinition of the fields, -- a
chiral Euclidean propagator. The pole of the propagator is due to a zero in
$\cos\beta$, i.e. a zero in the overlap (A.9).

Using the expansions (A.7$'$) one can express the perturbation $V$ in terms of
creation and annihilation operators,
\begin{eqnarray*}
V &=& \sum_{12}\Bigl[ b(1)^\dagger u(1)^\dagger V(1,2)u(2)b(2)+b(1)^\dagger
u(1)^\dagger V(1,2)v(2)d(2)^\dagger\Bigr.\\
&&\Bigl. +d(1)v(1)^\dagger V(1,2)u(2)b(2)+d(1)v(1)^\dagger
V(1,2)v(2)d(2)^\dagger\Bigr]
\end{eqnarray*}
where the symbol $\mathop{\sum}\limits_{12}$ denotes integration over $p_1,p_2$
and summation over $\sigma_1,\sigma_2$. The creation and annihilation operators
$b(1)^\dagger, d(1)$ together with their associated eigenspinors $u(1)^\dagger,
v(1)^\dagger$ may be of the $+$ or $-$ type. Likewise for $b(2), d(2)^\dagger$,
etc. The operator $V$ either creates pairs or scatter. In second order one
finds either 2-- or 4--particle intermediate states. The only unusual feature
of the overlap computations is the appearance of matrix elements like
\begin{eqnarray*}
\langle 1,\bar 2+\vert -\rangle &=& \langle +\vert d_+(2)b_+(1)\vert -\rangle\\
&=& \langle +\vert d_+(2)d_-(1)^\dagger \vert -\rangle \sin\beta (1)\\
&=&\delta (1,2)\ \langle +\vert -\rangle\ \tan\beta (1)
\end{eqnarray*}
where $\delta (1,2)$ stands for $(2\pi
)^4\delta_{2\pi}(p_1-p_2)\delta_{\sigma_1\sigma_2}$. Here we have used (A.8) to
express $b_+(1)$ in terms of $b_-(1)$ and $d_-(1)^\dagger$. The ground state
$\vert -\rangle$ is annihilated by $b_-(1)$. In passing to the next line we
have used the formulae (A.8$'$), (A.9) together with the Fock condition
$d^\dagger_\pm\vert 0\rangle =0$. In commuting $d_-(1)^\dagger$ to the right it
creates a gap at $k_1,\sigma_1$ in the product over $d_-$ that defines the
ground state $\vert -\rangle $. Likewise, in commuting $d_+(2)$ to the left it
creates a gap at $k_2,\sigma_2$ in the product over $D^\dagger_+$ that defines
$\langle +\vert$. The resulting states are orthogonal unless $1=2$ in which
case their overlap equals $\langle +\vert -\rangle /\cos\beta (1)$. Quite
generally, any overlap matrix element can be reduced with the help of (A.*) to
a linear combination of terms like
\begin{eqnarray}
&&\langle +\vert d_+(i_1)\dots d_+(i_n) d^\dagger_-(j_n)\dots
d^\dagger_-(j_1)\vert -\rangle =\nonumber\\
&&\hspace{3cm} \langle +\vert -\rangle (\cos\beta (i_1)\dots\cos\beta
(i_n))^{-1}\sum_\pi\ \varepsilon_\pi\ \delta_{i_1,j_{\pi_1}}\dots
\delta_{i_nj_{\pi_n}}
\end{eqnarray}
where $\pi_1,\dots ,\pi_n$ is a permutation of $1,\dots ,n$.

\newpage

\section*{Appendix B -- Ward identities}

Gauge transformations of the fermion fields can be implemented by unitary
operators,
\setcounter{equation}{0}
\renewcommand{\theequation}{B.\arabic{equation}}
\begin{equation}
\psi (n)\to e^{i\theta (n)}\ \psi (n) =e^{-iF_\theta}\ \psi (n)\ e^{iF_\theta}
\end{equation}
where $\theta (n)$ is an hermitian matrix and
\begin{equation}
F_\theta =\sum_n\ \psi (n)^\dagger\ \theta (n)\ \psi (n)
\end{equation}
The Hamiltonians (2.1) are covariant in the sense
\begin{equation}
e^{iF_\theta}\ H_\pm (A,\phi )\ e^{-iF_\theta} =H_\pm (A^\theta ,\phi^\theta )
\end{equation}
where the transformed background fields $A^\theta$ and $\phi^\theta$ are
defined
by
\begin{eqnarray}
e^{i\theta (n)}\ \phi (n)\cdot T\ e^{-i\theta (n)} &=& \phi^\theta (n)\cdot
T\nonumber \\
e^{i\theta (n)}\ U(n,m\vert A)\ e^{-i\theta (m)} &=& U(n,m\vert A^\theta )
\end{eqnarray}
Since the functional $U(n,m\vert A)$ is defined by the path--ordered
exponential (2.4) it follows that $A^\theta (x)$ is given by
\begin{equation}
A^\theta_\mu (x) =e^{i\theta (x)}(A_\mu (x) +i\partial_\mu ) \ e^{-i\theta (x)}
\end{equation}
where $\theta (x)$ is a smooth function that interpolates the lattice values,
$\theta (n)$.

With Hamiltonians transforming according to (B.3) it follows that the
corresponding non--degenerate ground states must transform such that
\begin{equation}
e^{iF_\theta}\vert A,\phi\pm\rangle =\vert A^\theta ,\phi^\theta\pm\rangle\
e^{i\Phi_\pm (\theta ,A,\phi )}
\end{equation}
where the angles $\Phi_\pm$ are {\em real}. The effective action (2.10)
therefore satisfies the identity
\begin{equation}
\Gamma (A^\theta ,\phi^\theta )=\Gamma (A,\phi )+i\Phi_+(\theta ,A,\phi
)-i\Phi_-(\theta ,A,\phi )
\end{equation}
Its real part is gauge invariant, but its imaginary part is not, unless
$\Phi_+=\Phi_-$. These angles can be computed perturbatively. To first order in
$\theta$, for example,
\begin{eqnarray}
\Phi_+ &=& {\rm Re}\ {\langle +\vert F_\theta\vert A,\phi +\rangle\over\langle
+\vert A,\phi +\rangle}\nonumber\\
&=&{\rm Re}\  \langle +\vert F_\theta (1+G_+V+G_+VG_+V+\dots )\vert +\rangle
\end{eqnarray}
where the Brillouin--Wigner phase convention, $\langle +\vert A,\phi +\rangle
>0$, is assumed. The difference, $\Phi_+-\Phi_-$, is a pseudoscalar with
respect to space reflections and it can be shown to vanish up to first order in
$A_\mu$ and $\partial_\mu\phi$ in dimensions $\geq 4$. In second order it can
be non--vanishing, thereby reflecting the usual anomalous breakdown of gauge
symmetry associated with chiral fermions.

The formula (B.7) comprises the Ward identities for the fermion vacuum
amplitude. For infinitesimal $\theta$ we have, from (B.4) and (B.5)
\begin{eqnarray}
\delta\ A_\mu (x) &=& \partial_\mu\theta (x) -i[A_\mu (x),\theta
(x)]\nonumber\\
&=& \nabla_\mu\theta (x)\\[3mm]
\delta\phi (x)\!\cdot\! T &=& i[\theta (x),\phi (x)\!\cdot\! T]\nonumber
\end{eqnarray}
where $\phi (x)$ is a multiplet of smooth fields that interpolates the lattice
values $\phi (n)$. For writing Ward identities it is convenient to choose a
hermitian basis with real components, $\phi^i(x)$,
\begin{eqnarray}
\phi\cdot T &=& \phi^i T_i\nonumber\\
{[}T_j ,T_\alpha ] &=& i(t_\alpha )_j\ ^i\ T_i
\end{eqnarray}
so that the components transform according to
\begin {equation}
\delta\phi^i(x) =-\theta^\alpha (x)\ \phi^j(x) \ (t_\alpha )_j\ ^i
\end{equation}
With this notation the identity (B.7) implies
\begin{equation}
\nabla_\mu\ {\delta\Gamma\over\delta A^\alpha_\mu} +\phi^j(t_\alpha )_j\ ^i\
{\delta\Gamma\over\delta \phi^i}=-i\ {\delta\over\delta\theta^\alpha}\
(\Phi_+-\Phi_-)
\end{equation}
and, in particular, for the second derivates evaluated at $A_\mu
=\partial_\mu\phi =0$,
\setcounter{equation}{0}
\renewcommand{\theequation}{B.13\alph{equation}}
\begin{eqnarray}
\partial_\mu\ {\delta^2\Gamma\over\delta A^\alpha_\mu (x)\ \delta A^\beta_\nu
(0)}&+&\phi^Tt_\alpha\ {\delta^2\Gamma\over\delta\phi (x)\ \delta A^\beta_\nu
(0)}=0\\
\partial_\mu\ {\delta^2\Gamma\over\delta A^\alpha_\mu (x)\
\delta\phi (0)}&+&\phi^Tt_\alpha\ {\delta^2\Gamma\over\delta\phi (x)\ \delta
\phi (0)}= -\delta_4(x)\ t_\alpha\ {\delta\Gamma\over\delta\phi}
\end{eqnarray}
since $\Phi_+=\Phi_-$ does not contribute here. The vacuum polarization tensor
is not transverse unless $\phi =0$. Its longitudinal part is related to the
Higgs field kinetic term. To see what is implied by this relation, consider the
leading terms in a continuum approximation,
$$
\Gamma (A,\phi )\simeq \int d^4x\left[ {Z_3\over 4} (F^\alpha_{\mu\nu}
)^2+{Z_2\over 2} (\nabla_\mu\phi^i)^2+\dots\right]  \eqno({\rm B}.14)
$$
where $Z_2$ and $Z_3$ are invariant functions of $\phi (x)$. For this
simplified functional we have
\setcounter{equation}{0}
\renewcommand{\theequation}{B.15\alph{equation}}
\begin{eqnarray}
\delta^2\Gamma\over\delta A^\alpha_\mu (x)\ \delta A^\beta_\nu (0) &=&
Z_3\ \delta_{\alpha\beta}(\delta_{\mu\nu}\partial^2-\partial_\mu\partial_\nu
)\delta_4(x)\nonumber\\
&&-Z_2\ \phi^Tt_\alpha t_\beta\phi\ \delta_{\mu\nu}\ \delta_4 (x)\\[3mm]
{\delta^2\Gamma\over\delta\phi (x)\ \delta A^\beta_\nu (0)} &=&
Z_2\ t_\beta\phi\ \partial_\nu\ \delta_4(x)
\end{eqnarray}
which clearly satisfies (B.13a).

\end{document}